\documentclass{article}
\usepackage{a4wide,amsmath,amsfonts,epsf,mncite}
\def\bildchen#1#2{\begin{minipage}{#1}\epsfxsize=#1\epsfbox{#2}\end{minipage}}

\def\ddx{{{\rm d}^d\!x\,}}
\def\ddy{{{\rm d}^d\!y\,}}
\def\bx{{\bf x}}
\def\by{{\bf y}}
\def\cG{{\cal G}}
\def\cI{{\cal I}}
\def\cO{{\cal O}}

\begin{document}

\title{Koenderink Filters and the \\Microwave Background$^{*}$}
\author{Jens Schmalzing$^{1,2,\dag}$}

\date{Schlo{\ss} Ringberg, October 16--19 1996}

\maketitle

\renewcommand{\thefootnote}{\fnsymbol{footnote}}
\footnotetext[1]{ Proc.\ $2^{\rm nd}$ SFB workshop on {\em
Astro--particle physics}, Ringberg 1996, Report SFB 375/P002 (1997),
Ralf Bender, Thomas Buchert, Peter Schneider (eds.), in press.}
\renewcommand{\thefootnote}{\arabic{footnote}}
\footnotetext[1]{Ludwig--Maximilians--Universit{\"a}t,
Theresienstra{\ss}e 37, 80333 M{\"u}nchen, Germany}
\footnotetext[2]{Max--Planck--Institut f{\"u}r Astrophysik,
Karl--Schwarzschild--Stra{\ss}e 1, 85740 Garching, Germany}
\renewcommand{\thefootnote}{\fnsymbol{footnote}}
\footnotetext[2]{email jensen@mpa-garching.mpg.de}
\renewcommand{\thefootnote}{\arabic{footnote}}

\begin{abstract}
We introduce Koenderink filters as novel tools for statistical
cosmology. Amongst several promising applications, they provide a test
for the Gaussianity of random fields. We focus on this application and
present some preliminary results from an analysis of the Cosmic
Microwave Background (CMB).
\end{abstract}

\setcounter{tocdepth}{3}

\section{Introduction}

The anisotropies in the Cosmic Microwave Background Radiation (CMBR)
are the oldest features of the Universe accessible to
observations. Coming directly from the last scattering surface, they
supply information on structure at early epochs. Therefore they
greatly help in the cosmologists' task to constrain and outrule
various models of structure formation.

However, the Microwave sky as observed today is not merely of cosmic
origin. Various sources of noise and other signals obscure and distort
our observations. It is thus essential to apply sophisticated data
analysis techniques in order to gain at least some insight into cosmic
evolution. To overview just a few of the many methods employed,
consult {}\scite{kogut:tests}, {}\scite{tegmark:angular}, or
{}\scite{ferreira:nongaussian}.

In this talk we present a novel statistical method that originated in
medical imaging. Considering the temperature anisotropies as a
realization $u_0(\bx)$ of a continuous random field, we systematically

\begin{itemize}
\item 
remove structure on specific scales, and thus enhance signal over
noise,
\item 
describe local geometry by a small yet complete set of measures,
\item 
construct meaningful and interpretable descriptors to assess
statistical properties.
\end{itemize}

Along these lines of thought, {}\scite{koenderink:structure} developed
an approach to the analysis of flat, two--dimensional images that is
now successfully implemented in computer vision and medical imaging
{}\cite{terhaar:introduction}.  We have generalized his approach to
fields both on curved supports and in higher dimensions. A number of
applications to cosmological datasets have been developed in
{}\scite{schmalzing:diplom}, we will now briefly outline the method
and then present some preliminary results from an analysis of the CMB.

\section{The Method of Koenderink Filters}

\begin{figure}
\begin{center}
\begin{minipage}{5cm}\begin{center}
\bildchen{5cm}{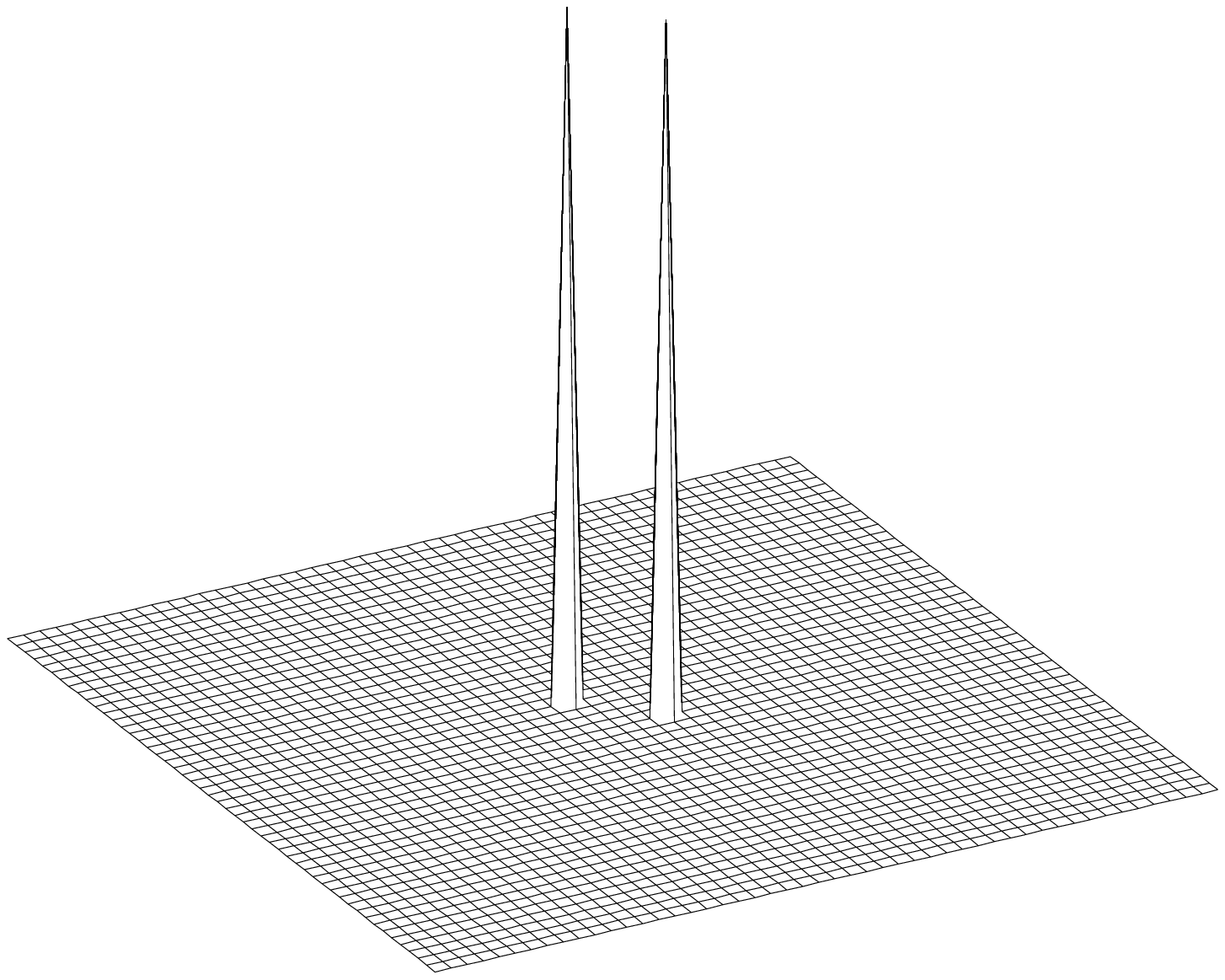} \\ $u_0$
\end{center}\end{minipage}
$\stackrel{\cG_\sigma}{\longrightarrow}$
\begin{minipage}{5cm}\begin{center}
\bildchen{5cm}{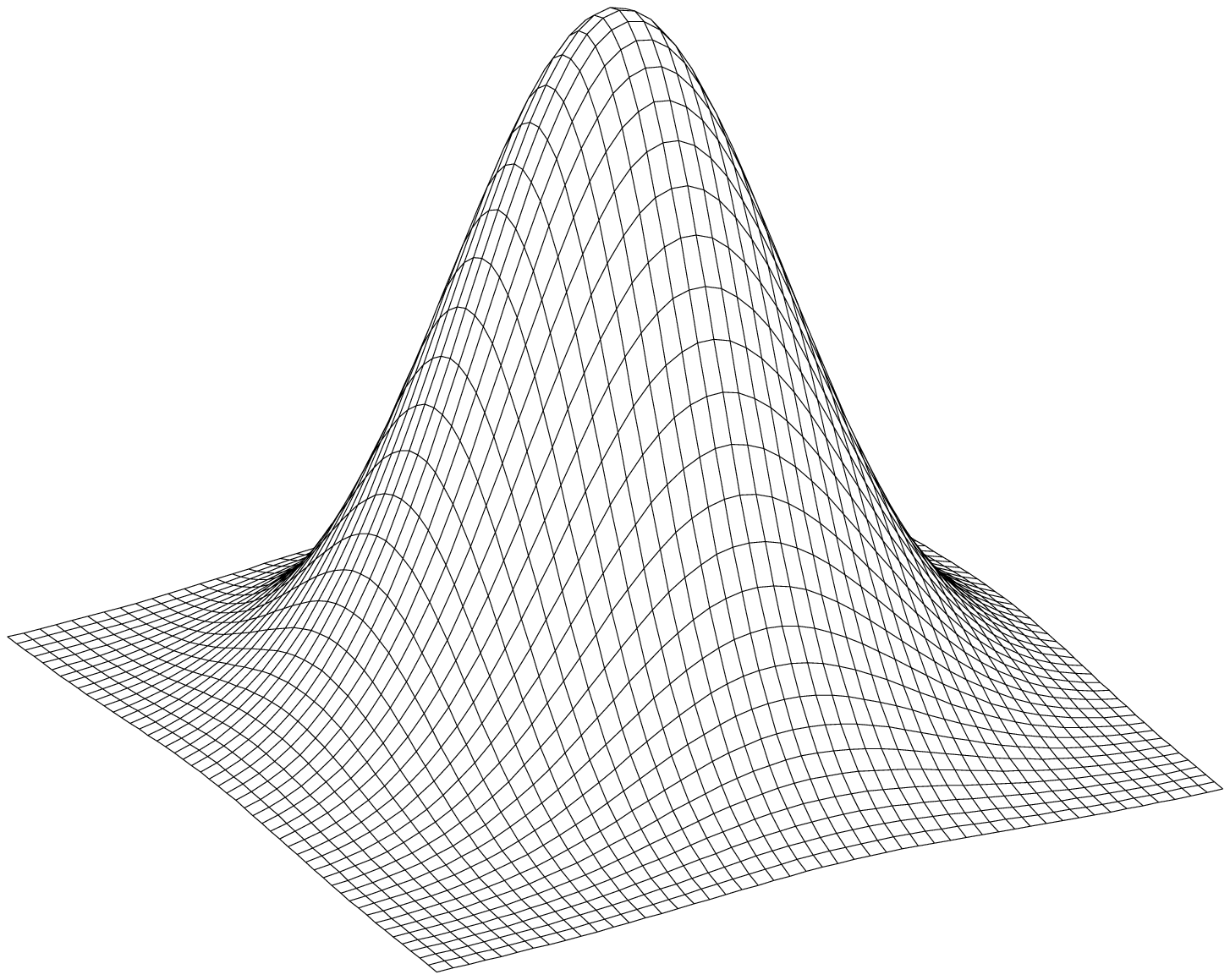} \\ $u_\sigma$
\end{center}\end{minipage}
\end{center}
\caption{
We demonstrate the idea behind the scale--space operator $\cG_\sigma$
described in the text. Structure on scales below $\sigma$ is removed
from the original field $u_0$, leaving a new field $u_\sigma$. In this
example, the separation of the two spikes disappears as filtering with
a larger scale merges them into one featureless blob.}
\end{figure}

\begin{figure}
\begin{center}
\bildchen{3.75cm}{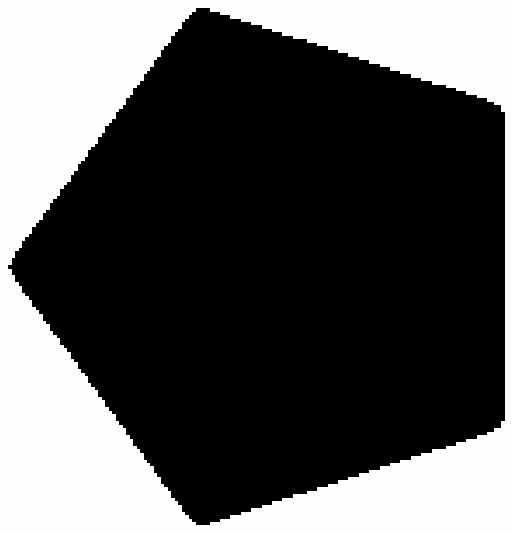}
\bildchen{3.75cm}{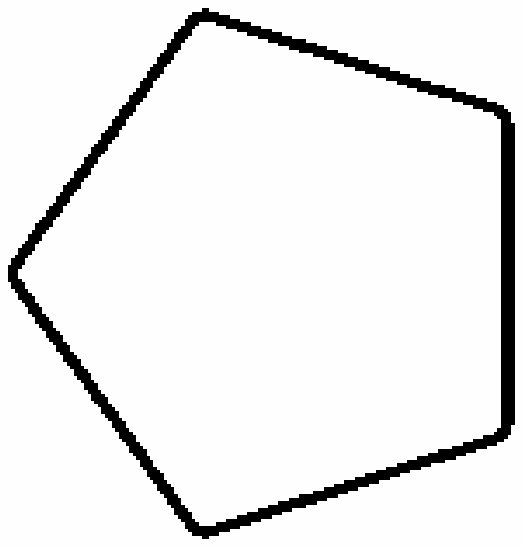}
\bildchen{3.75cm}{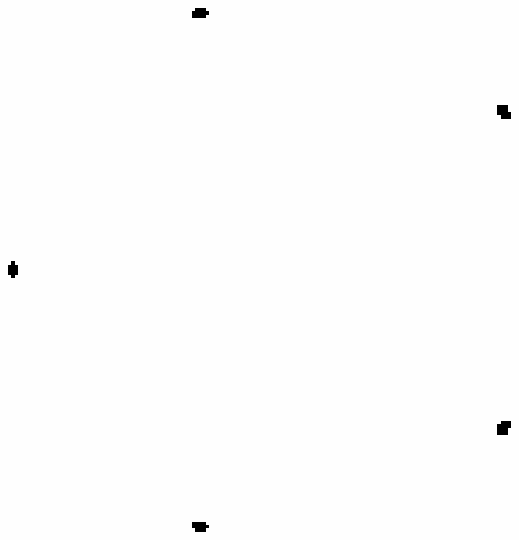} \\
\bildchen{3.75cm}{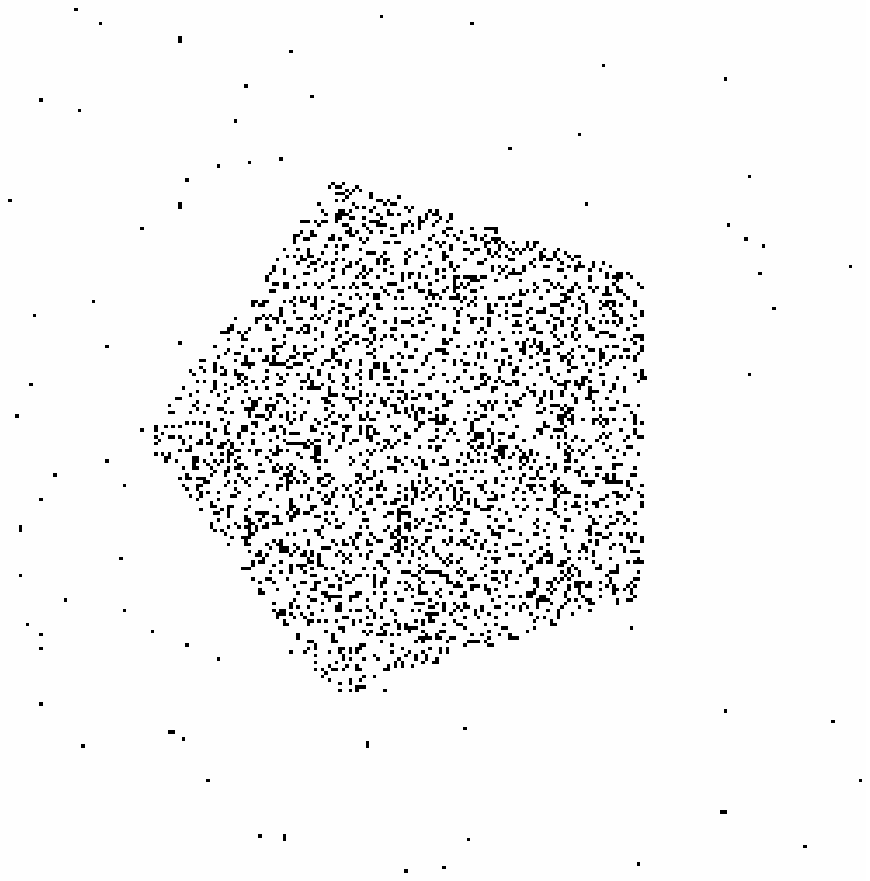}
\bildchen{3.75cm}{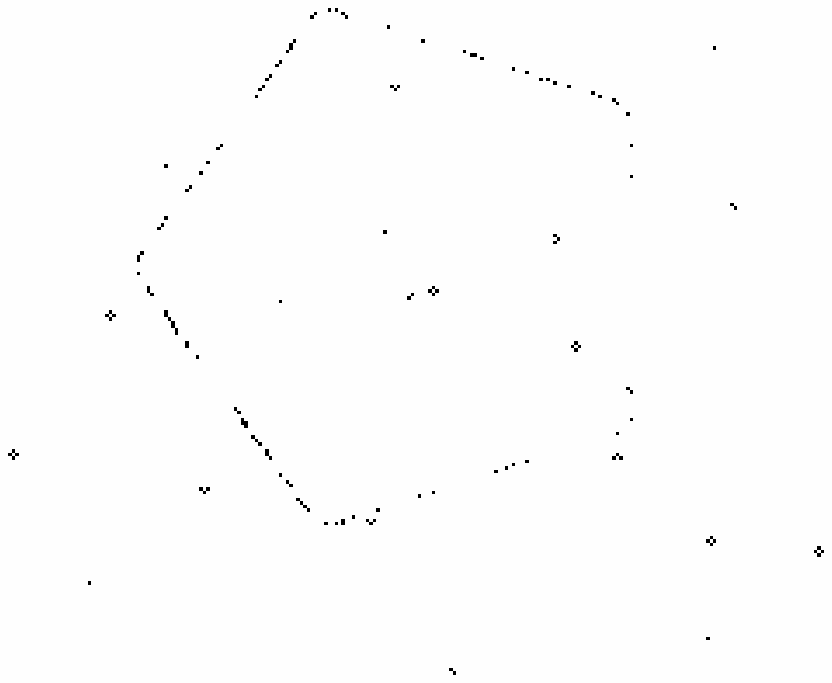}
\bildchen{3.75cm}{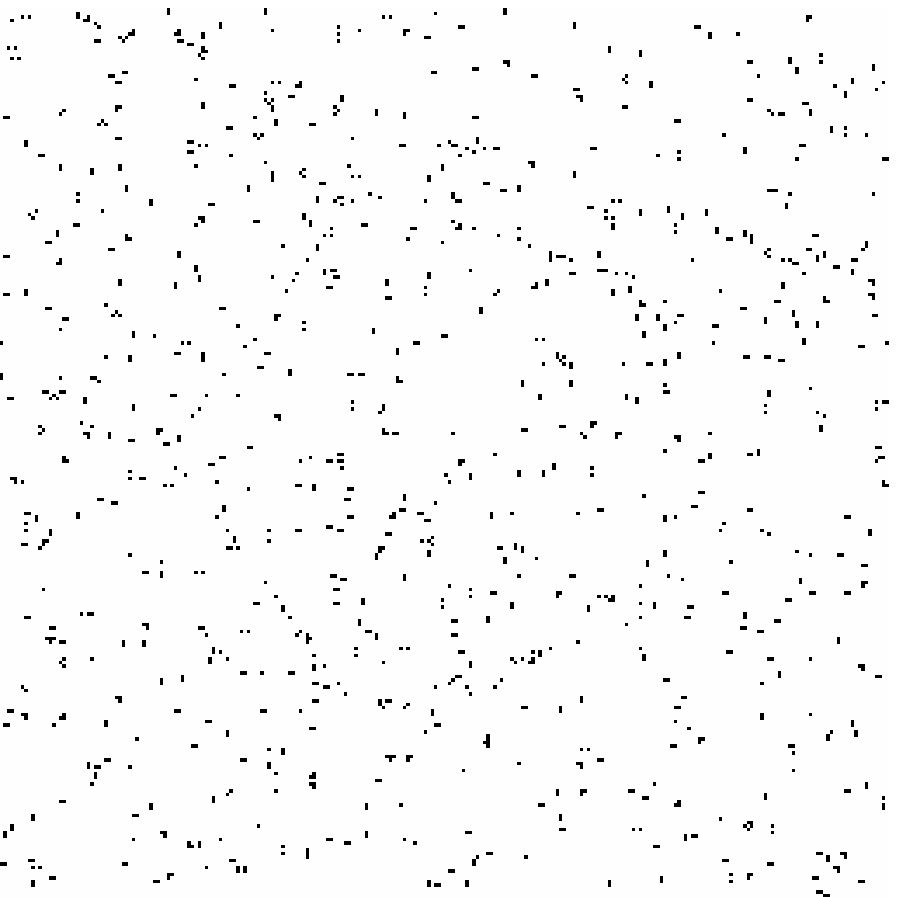} \\
\bildchen{3.75cm}{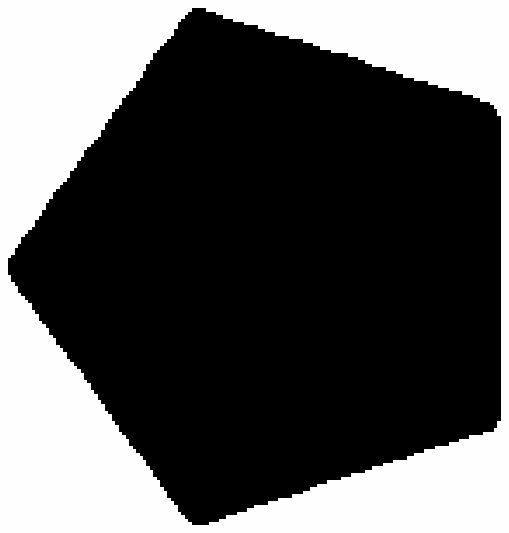}
\bildchen{3.75cm}{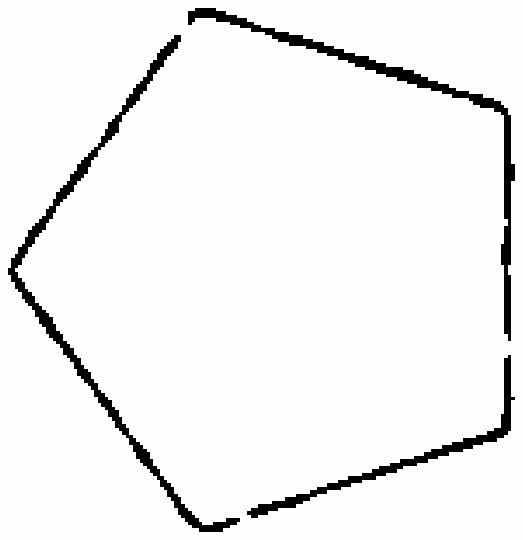}
\bildchen{3.75cm}{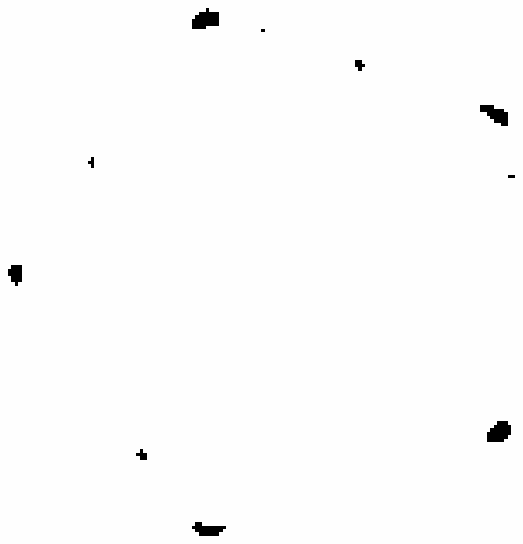}
\end{center}
\caption{\label{pentagon} 
These panels illustrate the application of scale--space operators (see
Section~\ref{scale--space}) and geometric invariants (see
Section~\ref{geometric}) to a two--dimensional image. All nine panels
were constructed from the image in the top left corner. The first row
shows a pentagon and two of the geometric invariants, namely edgeness
and cornerness which obviously enhance the corresponding features. In
the second row we see the pentagon disturbed by Poissonian noise
points of equal strength. Although the pattern is still visible, the
geometric invariants -- again we show cornerness in the middle column
and edgeness in the right column -- fail to find the inherent
features. However, by applying a scale--space filter of moderate
width, we arrive at the series of images shown in the bottom row. Both
the original pentagon and its edges and corners as seen by geometric
invariants have been reconstructed, although the noise is still
present.}
\end{figure}

\subsection{Scale--Space Operators}
\label{scale--space}

In order to separate the various scales present in a given field $u_0$
we wish to erase structure on scales below a given length $\sigma$,
thus arriving at a new field $u_\sigma$. In the most general approach,
this is done by applying an operator $\cG_\sigma$, that is
\begin{equation}
u_\sigma = \cG_\sigma u_0.
\end{equation}
The precise form of the so--called scale--space operator remains to be
specified.

It has been shown by {}\scite{koenderink:geometry} (see also the
appendix of {}\pcite{koenderink:scalespace}) that the natural choice
for $\cG_\sigma$ is the convolution with a Gaussian kernel of width
$\sigma$:
\begin{equation}
G_\sigma(\bx,\by) = \frac{1}{(2\pi\sigma^2)^{d/2}}
\exp\left(\frac{||\bx-\by||^2}{2\sigma^2}\right).
\end{equation}

In order to achieve this uniqueness we require the following simple
and actually fairly compelling properties of the operator:
\begin{itemize}
\item{\bf Additivity and Closure:} 
Iterating two scale--space operators is equivalent to applying a single
scale--space operator with an ``added'' scale; it turns out that only
with additivity of the squares, that is $\sigma_1\oplus\sigma_2 =
\sqrt{\sigma_1^2+\sigma_2^2}$ we can satisfy all requirements
simultaneously.
\begin{equation}
\cG_{\sigma_1}\circ\cG_{\sigma_2}=\cG_{\sigma_1\oplus\sigma_2}.
\end{equation}
\item{\bf Limits:} 
Erasing no structure leaves the image unaltered, while erasing
structure on all scales erases the image completely.
\begin{equation}
\cG_{\sigma\rightarrow{0}}={\bf 1}\mbox{ and }
\cG_{\sigma\rightarrow\infty}={\bf 0}.
\end{equation}
\item{\bf Linearity:} 
We assume validity of the superposition principle, that is structure
on one scale is not influenced by features on another scale. The
operator is thus linear and can be expressed as a convolution with an
appropriate kernel,
\begin{equation}
u_\sigma(\bx) = \int\ddx\int\ddy G_\sigma(\bx,\by)u_0(\by).
\end{equation}
\item{\bf Invariance:} 
Rotational and translational invariance reduce the dependence of the
kernel on two points to a dependence on a single scalar, the distance
of the points.
\begin{equation}
G_\sigma(\bx,\by) = G_\sigma(||\bx-\by||).
\end{equation}
\end{itemize}

\subsection{Geometric Invariants}
\label{geometric}

\begin{table}
\begin{center}
\begin{tabular}{ll}
$u_{,i}u_{,i}                            $ & edgeness \\
$u_{,ii}                                 $ & profile's extrema \ldots \\
$u_{,ii}u_{,jj}-u_{,ij}u_{,ij}           $ & \ldots and saddlepoints \\
$u_{,i}u_{,ij}u_{,j}-u_{,i}u_{,i}u_{,jj} $ & cornerness
\end{tabular}
\end{center}
\caption{\label{basis}
This table summarizes a choice of basic invariants that is unique and
complete to second order in two dimensions. Furthermore, all
invariants may be interpreted as morphometric quantities as given in
the right column.}
\end{table}

The simplest way to characterize local geometry around a point of the
support of a random field is to expand the field as a Taylor series to
sufficient order $N$ around this point. The first two orders, and the
general expression are\footnote{We use the summation convention for
pairwise indices in a product, and denote partial derivatives by
indices following a comma.}
\begin{equation}
\begin{split}
u_\sigma(\bx+\by)&=u_{\sigma,i}(\bx)y_i+\cO(y^2)\\
u_\sigma(\bx+\by)&=u_{\sigma,i}(\bx)y_i+\frac{1}{2}
u_{\sigma,ij}(\bx)y_iy_j+\cO(y^3)\\
&\vdots \\
u_\sigma(\bx+\by)&= \sum_{M=0}^N \frac{u_{\sigma,i_1{\ldots}i_M}(\bx)}{M!}
y_{i_1}{\ldots}y_{i_M} + \cO(y^{N+1}). 
\end{split}
\end{equation} 
Obviously, considering the derivatives up to $N$-th order allows us to
gain insight into the field's local behaviour while discarding higher
orders in a controlled manner. However, mere derivatives depend on the
choice of coordinate system. If we remove these unwanted degrees of
freedom by considering coordinate independent invariants such as the
square of the gradient instead of its components, we arrive at a
description of local geometry that is both complete, in the sense that
it contains all information up to $N$-th order, and unique, in the
sense that all invariants can be constructed as polynomials in a few
basic invariants.

As an important example we will consider a basis for invariants up to
second order in two dimensions. By counting degrees of freedom --
derivatives of first and second order give five independent numbers,
but one of these degrees of freedom is removed when fixing the
coordinate system -- we conclude that four invariants, apart from the
field $u$ itself, can be found. A simple choice is
\begin{equation}
u_{,i}u_{,i}\quad u_{,ii}\quad u_{,i}u_{,ij}u_{,j}\quad u_{,ij}u_{,ji}
\end{equation}

Not all of these basic invariants allow for an intuitive
interpretation. We have to go to a different, equivalent basis which
is summarized in Table~\ref{basis}, together with geometric
interpretations. $u_{,i}u_{,i}$ is the square of the gradient and is
expected to become large at the steepest edges, hence the term
``edgeness''. The quantities $u_{,ii}$ and
$u_{,ii}u_{,jj}-u_{,ij}u_{,ji}$ are nothing but the trace and
determinant of the Hessian matrix of the field's profile and thus
provide information about the nature of extrema. Finally,
$u_{,i}u_{,ij}u_{,j}-u_{,i}u_{,i}u_{,jj}$ is related to the curvature
of isolines, which becomes large at corners of the field's isolines
and was accordingly named ``cornerness'' by
{}\scite{koenderink:scalespace}.

In Figure~\ref{pentagon} we demonstrate the effect of scale--space
operators and geometric invariants on a simple pattern.

\subsection{Statistical Descriptors}
\label{sec:descript}

After applying scale--space filters and calculating geometric
invariants, we arrive at a field $\cI_\sigma$ that contains less
unwanted structure for an appropriate choice of $\sigma$ and enhances
geometric features according to the invariant that has been chosen.
Several possibilities exist to construct statistical descriptors from
these fields.

\subsubsection{Isodensity contours} 

The surface integration necessary for the calculation of the genus of
isodensity contours {}\cite{gott:sponge} can be reduced to taking a
spatial mean value of Koenderink invariants, multiplied with a delta
function for selecting the specific isodensity contour. For the Euler
characteristic $\chi=1-g$ of the isodensity contour of the field
$u(\bx)$ to the threshold $\nu$, for example, one obtains
\begin{equation}
\chi(\nu) = \frac{1}{2\pi}
\int\!\ddx\frac{u_{,i}u_{,ij}u_{,j}-u_{,i}u_{,i}u_{,jj}}{u_{,k}u_{,k}}
\delta(u-\nu)
\end{equation}
Recently, this rationale has been used by {}\scite{schmalzing:beyond}
to calculate not only the genus, but all Minkowski functionals for a
continuous random field in three dimensions.

\subsubsection{Excursion sets} 

It is possible to measure the size of an invariant's excursion set
over its mean value -- the sharper the features emphasized by the
invariant, the smaller the excursion set.  Double Poisson processes
for various geometrical objects can be efficiently discriminated by
measuring the excursion sets and successively filtering out larger and
larger scales in the point distribution (see
{}\pcite{schmalzing:diplom} for further details).

\subsubsection{Probability density} 

Finally, one can think of using the probability density of an
invariant as an indicator of certain patterns. It is especially
promising to compare the measured curves to analytically calculated
mean values for simple types of random fields.  We have developed a
test for the Gaussianity of random fields and will present preliminary
results of its application to CMB anisotropies in the following
Section~\ref{koe.gauss}.

%+---------------------+
%|     Gaussianity     |
%+---------------------+
\section{Gaussianity of Random Fields}
\label{koe.gauss}

It is generally believed that the Cosmic Microwave Background
anisotropies obey the statistics of a Gaussian random field, at least
on scales observed by the COBE satellite.  On smaller scales the
situation is not so clear. In any case it is important to assess
whether there are any non--Gaussian signatures in the microwave sky.

\begin{figure}
\begin{center}
\bildchen{8cm}{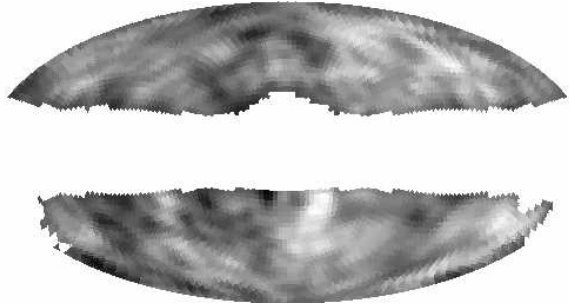}
\end{center}
\caption{
This figure shows a Microwave sky map constructed from the COBE DMR
four--year data. By removing the monopole and dipole from the mean
value $(A{+}B)/2$ of the 53GHZ channels, the anisotropies on smaller
scales become visible. Finally, we only use the 3881 pixels not
covered by COBE group's mask for galactic emission to plot this map
and to apply Koenderink invariants.}
\end{figure}

\begin{figure}
\begin{center}
\bildchen{14cm}{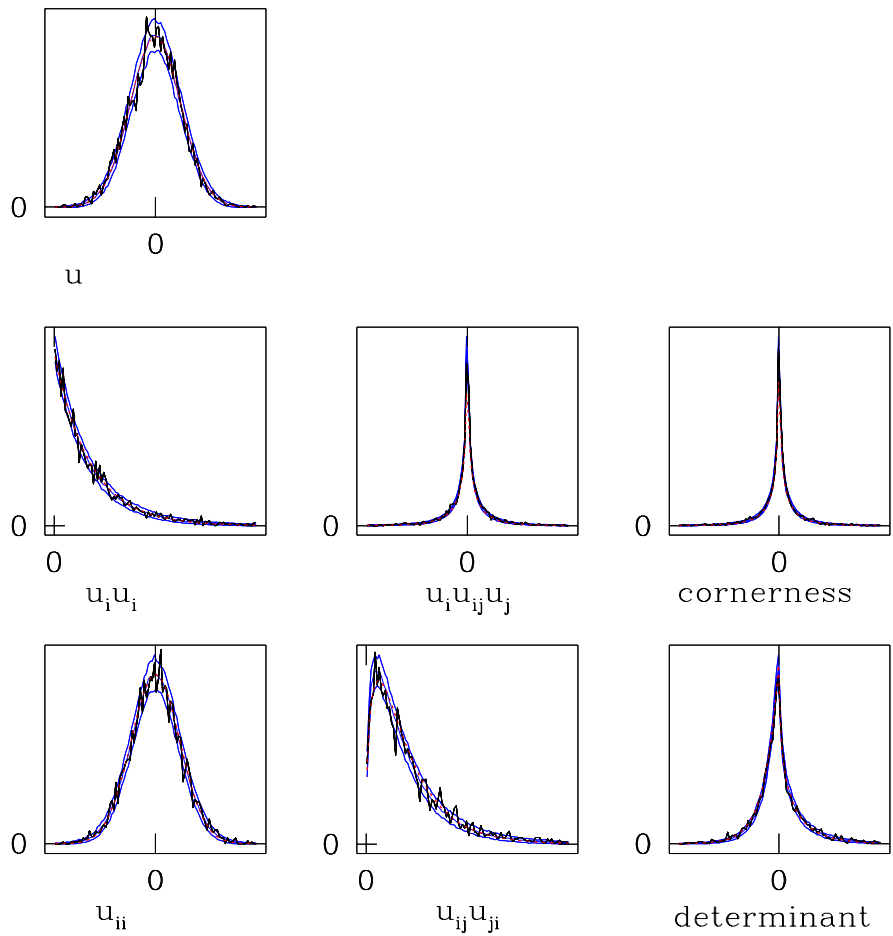}
\end{center}
\caption{\label{koe.cmbr.53}
Here we display the results of an analysis of the map in the previous
figure. The probability densities of several Koenderink invariants are
shown, both as measured from the data (solid line) and as calculated
for 100 realizations of a Gaussian random field (the shaded area
denotes $1\sigma$ fluctuations).  }
\end{figure}

\begin{figure}
\begin{center}
\bildchen{8cm}{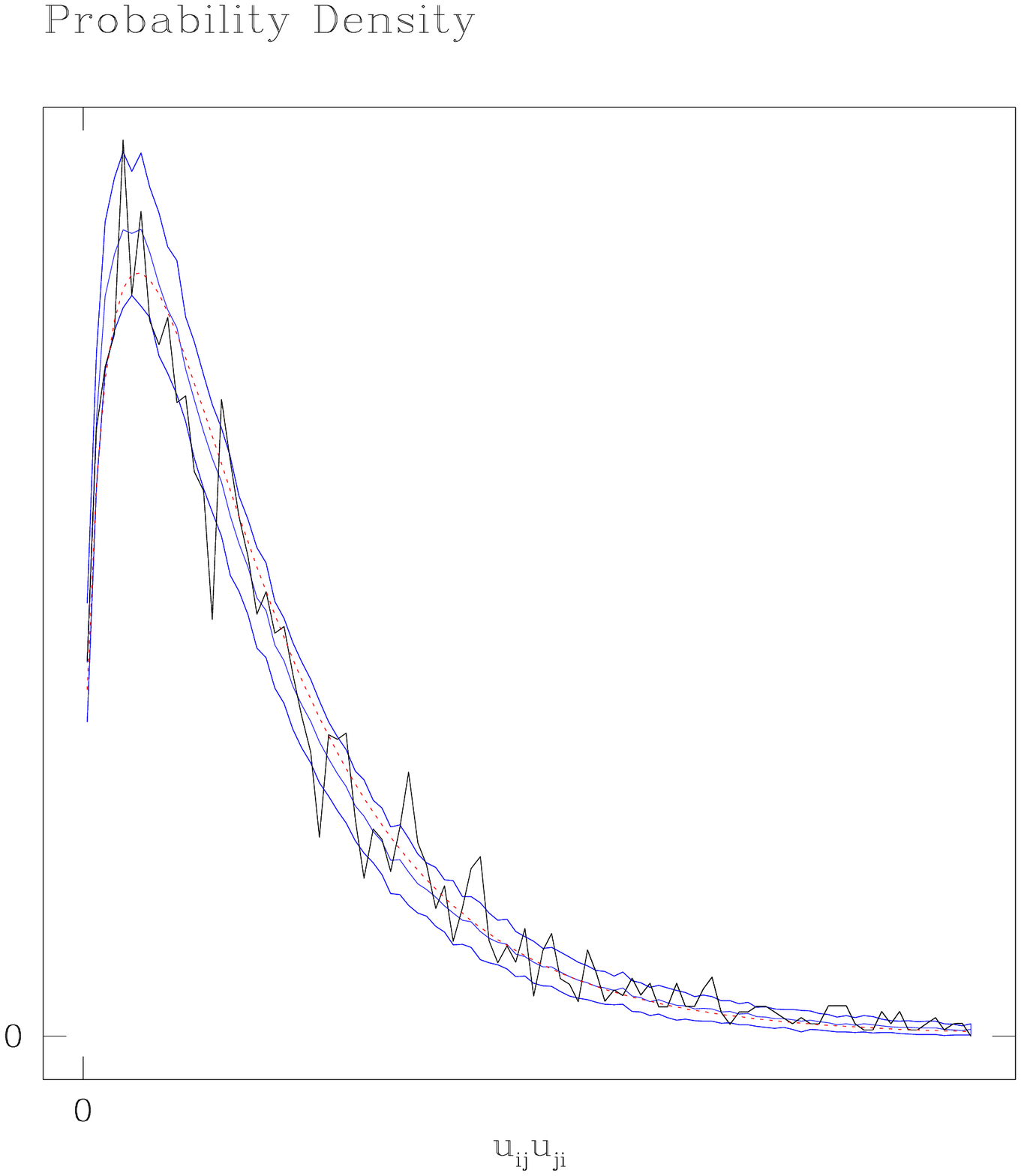}
\end{center}
\caption{\label{koe.cmbr.53.invariant4}
This plot shows the probability density of one of the invariants from
Figure~\protect\ref{koe.cmbr.53}. The solid, irregular line was
measured from the COBE DMR data. The area indicated by the three
smoother solid lines corresponds to $1\sigma$--fluctuations calculated
from one hundred realizations of a Gaussian random field.  Obviously
the fluctuations of the data around the fitted mean value (dotted
line) are consistent with a Gaussian random field. However, we do see
slight discrepancies; future work will assess whether they become
significant at higher resolution, when improved statistics reduces the
sample--to--sample variance. }
\end{figure}

\subsection{Gaussian Random Fields}

The statistical properties of homogeneous and isotropic Gaussian
random fields are solely determined by their power spectrum $P(k)$ or
equivalently their two--point correlation function $\xi(r)$. Among
other works, the standard textbook by {}\scite{adler:randomfields} or
the article by {}\scite{bardeen:gauss} are devoted to an extensive
study of random fields in general and Gaussian ones in particular; the
calculational schemes presented there also allow to tackle Koenderink
invariants of random fields analytically.

\subsection{Koenderink Invariants}

It turns out that the probability densities of any Koenderink invariant
(see {}\pcite{schmalzing:diplom} for detailed results) depends solely
on three common parameters ($\xi(0)$, $\xi''(0)$ and
$\xi''''(0)$). Furthermore, they only determine the width of the
distribution, while the shape is the same for any Gaussian random
field.  They can be fitted using the functions measured from the
random field, and the goodness of the fit gives a measure of the
deviation from Gaussianity.

The panels shown in Figure~\ref{koe.cmbr.53} illustrate the method
outlined above. We show a random field, the COBE $(A{+}B)/2$ signal in
the 53GHz channel with a customized galactic cut (details can be found
in {}\pcite{bennett:fouryeardata}). From this random field several
distribution functions of Koenderink invariants have been
calculated. Although the behaviour is largely consistent with the
assumption of a Gaussian random field, some deviations are
visible. Figure~\ref{koe.cmbr.53.invariant4} shows and explains one of
the seven panels in more detail.

\section{Outlook}

Apart from the example discussed in this talk, Koenderink invariants
suggest a variety of further studies in cosmology.

\subsection{Large Scale Structure}

The possibility of extending Koenderink invariants to arbitrary
dimensions has received only brief attention in this article. On the
one hand, Koenderink filters can be used to construct morphological
statistics for the three--dimensional matter distribution in the
Universe {}\cite{schmalzing:diplom}, on the other hand, the insights
gained from the theoretical foundations may be used to shed new light
on seemingly unrelated measures such as genus statistics
{}\cite{schmalzing:beyond}.

\subsection{Microwave Background}

The application discussed in this talk uses Koenderink invariants to
assess whether non--Gaussian features can be seen in the Cosmic
Microwave Background. The preliminary results suggest that this is not
the case, at least for the scale probed by the COBE
satellite. However, a number of issues remain to be addressed.

Numerical and, if possible, analytical calculations for certain
non--Gaussian fields can test the significance of our findings and
test the discriminative power in comparison to other methods (e.g.\
{}\pcite{kogut:gaussian}).

With regards to the forthcoming high--resolution surveys of the
microwave sky {}\cite{bersanelli:cobrassamba,bennett:map}, the
performance of Koenderink filters on smaller scales needs to be
tested. Their noise reduction abilities are another important subject
of study.

Finally, it would be interesting to apply related descriptors other
than the probability densities of Koenderink invariants. The methods
outlined in Section~\ref{sec:descript} are promising candidates.

\addcontentsline{toc}{section}{Acknowledgements}
\section*{Acknowledgements}

I wish to thank R\"udiger Kneissel for many fruitful discussions and
comments. The workshop at Schlo{\ss} Ringberg was supported by the
Sonderforschungsbereich 375 f\"ur Astroteilchenphysik der Deutschen
Forschungsgemeinschaft. The COBE datasets were developed by the NASA
Goddard Space Flight Center under the guidance of the COBE Science
working Group and were provided by the NSSDC.

\addcontentsline{toc}{section}{Bibliography}
\providecommand{\bysame}{\leavevmode\hbox to3em{\hrulefill}\thinspace}

\end{document}